\documentclass[a4paper,11pt]{article}
\usepackage[numbers,sort&compress]{natbib}
\usepackage[all]{xy}
\usepackage{amsmath,amsfonts,amssymb,amsthm,epsfig,amscd,comment,latexsym,psfrag}
\usepackage{CJK}
\usepackage{mathrsfs}
\usepackage{nicefrac,xspace,tikz}
\usepackage{arydshln}
\usetikzlibrary{arrows}
\usepackage{graphicx}
\usepackage{caption,subcaption}
\usepackage{xcolor,float}
\usepackage{appendix}
\usepackage{graphicx}
\usepackage{caption}
\makeatletter

\newcommand{\Rmnum}[1]{\expandafter\@slowromancap\romannumeral #1@}
\makeatother

\def\footnoterule{\kern 1mm \hrule width 7cm \kern 2.2mm}%

\def\dsum{\displaystyle\sum}

\def\Tr{\mathrm{Tr}}

\newcommand{\ad}{\operatorname{ad}}

\allowdisplaybreaks
\begin{document}

\title{\vspace{1.5cm}\bf
	$W$-representations for multi-character partition functions and their $\beta$-deformations}


\author{ Lu-Yao Wang$^{a,}$\footnote{wangly100@outlook.com},
V. Mishnyakov$^{b,c,d}$\footnote{mishnyakovvv@gmail.com},
A. Popolitov$^{b,d,e,}$\footnote{popolit@gmail.com},
Fan Liu$^{a,}$\footnote{liufan-math@cnu.edu.cn},
Rui Wang$^{f,}$\footnote{wangrui@cumtb.edu.cn}
}
\date{}

\maketitle

\vspace{-6.5cm}

\begin{center}
	\hfill FIAN/TD-04/23\\
	\hfill IITP/TH-04/23\\
	\hfill ITEP/TH-05/23\\
	\hfill MIPT/TH-04/23
\end{center}

\vspace{3.5cm}

\begin{center}
$^a${\em School of Mathematical Sciences, Capital Normal University,
	Beijing 100048, China} \\
$^b$ {\em MIPT, Dolgoprudny, 141701, Russia}\\
$^c$ {\em Lebedev Physics Institute, Moscow 119991, Russia}\\
$^d$ {\em NRC ``Kurchatov Institute'', Moscow  123182, Russia}\\
$^e$ {\em Institute for Information Transmission Problems, Moscow 127994, Russia}\\
$^f${\em Department of Mathematics, China University of Mining and Technology,
	Beijing 100083, China}
\end{center}

\begin{abstract}
In this letter we continue the development of $W$-representations. We propose several generalizations of the known models, such as the hypergeometric Hurwitz $\tau$-functions. We construct $W$-representations for multi-character expansions, which involve a generic number of sets of time variables. We propose integral representations for such kind of partition functions which are given by tensor models and multi-matrix models with multi-trace couplings. We further propose the $\beta$-deformation of the discussed $W$-representation for the Hurwitz case for two sets of times as well as for the multi-character case.
\end{abstract}

\section{Introduction}
Recently a lot of progress has been made in understanding the connection between $W$-representations for $\tau$-functions and superintegrability of matrix models. In particular two main progressions have been made.

Firstly, a rather comprehensive construction of explicit $W$-operators, that generate a very large class of $\tau$-functions has been obtained. The main idea of this construction is the use of iterated commutator in the $\mathcal{W}_{\infty}$ algebra to construct families of commutative operators $W^{(m)}_{n} $ which are related to $p_n$ and $\frac{\partial}{\partial p_n}$ by an algebra automorphism.

Second is the matrix model realization of the discussed $\tau$-functions. The two matrix model and it's multimatrix generalization realise the character expansions generated by the $W$ operators. In particular, the coefficients of the character expansions are given by the matrix averages of characters and, hence, the property of superintegraibility of these matrix models is crucial.
\\\\
Here we would like to propose a few generalizations of the relation between $W$-operators, character expansions and matrix models. First is the generalization to multicharacter expansions. Such expansions would involve an arbitrary number of sets of time variables in contrast to the two sets of Toda $\tau$-functions. It turns out that the generalisation is straightforward on the level of $W$-operators. We propose the question of finding integral representations for such partition functions and outline the first steps in this direction. Such multicharacter expansions naturally appear in tensor models. The simplest example is provided by the gaussian tensor model of \cite{Itoyama2020}. Hence, in our discussion we will provide a $W$-representation, for this tensor model, which was given only implicitly in \cite{Itoyama2020}. Another option, that we will explore are the multimatrix models with multitrace interaction vertices. Using ideas from \cite{MMMP} \emph{we provide an explicit multi-matrix realisation for a rather general multi-character expansion}.

The other generalization is the so-called $\beta$-deformation. It has already been noticed in \cite{wangruiliufan,Bawane:2022cpd,Mishnyakov:2022bkg,Morozov:2019gbt}, that $W$-operators can be $\beta$-deformed. The main ingredient is the deformed operator $\mathcal W_0$, which is nothing but the Calogero-Sutherland hamiltonian:
\begin{eqnarray}\label{hurwitz-ope}
\mathcal{W}_0=\frac{1}{2}\sum_{k,l=1}^{\infty}\big(\beta(k+l)p_{k}p_l
\frac{\partial}{\partial p_{k+l}}+klp_{k+l}\frac{\partial}{\partial p_k}\frac{\partial}{\partial p_l}\big)
+\frac{1}{2}\sum_{k=1}^{\infty}((1-\beta)(k-1)+2\beta N)kp_{k}\frac{\partial}{\partial p_{k}},
\end{eqnarray}
As we will demonstrate, the deformation on the side of $W$-operators is once-again straightforward. Combined with the results of \cite{MMMP} this hint at the existence of the corresponding deformation of the multimatrix models and their multicharacter generalization.
\\\\
The paper is structured as follows. In section \ref{sec:multichar} we discuss the multicharacter expansions and their integral representations. In section \ref{sec:beta} we disucss the $\beta$-deformation.

\section{Multi-character expansion}\label{sec:multichar}
Let us briefly remind the construction of $W$-operators from \cite{MMMP}. As a starting point take:
\begin{equation}
   \hat W_0(u)= \dfrac{1}{2} \sum_{a, b=0}\left( (a+b) p_a p_b \frac{\partial}{\partial p_{a+b}}+a b p_{a+b} \frac{\partial^2}{\partial p_a \partial p_b}\right) + u\sum_{a} a p_a \dfrac{\partial }{\partial p_a},
\end{equation}
and
\begin{equation}
\begin{split}
    E_n(\vec{u}) &= \prod_{i=1}^{n} \ad_{\hat W_0(u_i)} p_1 = [\hat W_0(u_n) , [ \ldots ,[\hat W_0(u_1),p_1] \ldots]],
    \\
    F_n(\vec{u}) &= \prod_{i=1}^{n} \ad_{\hat W_0(u_i)} \dfrac{\partial}{\partial p_1} = \left[\hat W_0(u_n) , \left[ \ldots ,\left[\hat W_0(u_1),\dfrac{\partial}{\partial p_1}\right] \ldots\right] \right].
\end{split}
\end{equation}
Then, for each $n$ one can construct a family of \emph{commuting} operators:
\begin{equation}\label{W+-}
\begin{split}
        \hat W_{-k}^{(n)}(\vec{u})&=\frac{1}{(k-1)!}\ad^{k-1}_{E_{n+1}(\vec{u})} E_{n}(\vec{u}),
        \\
        \hat W_{k}^{(n)}(\vec{u})&=\frac{(-1)^{k-1}}{(k-1)!}\ad^{k-1}_{F_{n+1}
        (\vec{u})} F_{n}(\vec{u}).
\end{split}
\end{equation}
These operators are related to the time variables via an algebra automorphism, given by the operator $\hat{O}(u)$ defined by it's action on Schur functions \cite{1405}
\begin{equation}
	\hat{O}(u) S_R = \left( \prod_{(i,j) \in R} (u+j-i) \right) S_R
\end{equation}
then
\begin{equation}
\begin{split}
		W^{(n)}_{-k}(\vec{u}) &= \left(\prod_{i=1}^{n} \hat{O}(u_i) \right) p_k  \left(\prod_{i=1}^{n} \hat{O}(u_i) \right)^{-1}
			\\
		W^{(n)}_{k}(\vec{u}) &= \left(\prod_{i=1}^{n} \hat{O}(u_i) \right)^{-1} \dfrac{\partial }{\partial p_k}  \left(\prod_{i=1}^{n} \hat{O}(u_i) \right)
\end{split}
\end{equation}
The $W$-representations in \cite{MMMP} where constructed as a counterpart to the Cauchy identity:
\begin{equation}
    \begin{split}
        \exp\left( \sum_{k=1}^{\infty} \dfrac{\bar{p}_k p_k}{k} \right) &= \sum_R S_R(p) S_R(\bar{p}),\\
        &\Downarrow \\
        \exp\left( \sum_{k=1}^{\infty} \dfrac{\bar{p}_k \hat W^{(m)}_{-k}[p]}{k} \right)\cdot 1 &= \sum_R \prod_{i=1}^{m} \dfrac{S_R\{p_k=u_i\}}{S_R\{\bar{p}_k=\delta_{k,1}\}} S_R(p) S_R(\bar{p}),
        \\
        \exp \left(\sum_{k=1}^{\infty} \frac{\bar{p}_k \hat W_k^{(m)}[p]}{k}\right) \cdot \exp \left(\sum_{l=1}^{\infty} \frac{p_l g_l}{l}\right) &= \sum_{R,Q}\prod_{i=1}^m\frac{S_R\{p_k=u_i\}S_Q\{p_k=\delta_{k,1}\}}
        {S_R\{p_k=\delta_{k,1}\}S_Q\{p_k=u_i\}}S_{R/Q}\{\bar p\}\\
        &\ \ \ \times S_Q\{p\}S_R\{g\},
    \end{split}
\end{equation}
where:
\begin{equation}
\dfrac{S_{R}\{p_k=u\}}{S_{R}\{p_k=\delta_{k,1}\}} = \prod_{(i,j)\in R} \left( u+j-i \right).
\end{equation}
Now, one can make use of the generalized Cauchy formula, which involves multiple sets of time variables $p_k^{(i)}$
\cite{Itoyama1909}:
\begin{eqnarray}\label{generalized-Cauchy}
\exp\left(\sum_{k=1}^{\infty}\frac{\prod_{i=1}^rp^{(i)}_{k}}{k} \right)
=\sum_{R_1,\cdots,R_r}
C_{R_1,\cdots,R_r}\prod_{i=1}^r S_{R_i}\{p^{(i)}\},
\end{eqnarray}
with the  coefficient is given by
\begin{equation}
    C_{R_1,\cdots,R_r}=\dsum_{\Delta\vdash m}
\frac{\prod_{i=1}^r\psi_{R_i}(\Delta)}{z_{\Delta}},
\end{equation}
where $\psi_{R_i}(\Delta)$ is the character of symmetric group $S_m$,
 $z_{\Delta}:=\prod_i m_i!\cdot i^{m_i}$ is the standard symmetry factor of the Young diagram $\Delta=\{1^{m_1},2^{m_2},\cdots\}$. Clearly, these coefficients are nonzero only if the sizes $|R_i|$ of all the partitions are equal.
 \\\\
 Now, due to commutativity of the $W$-operators \eqref{W+-} we can freely substitute those operators instead some of the sets of time variables in the generalized Cauchy formula. Suppose we have $r=r_1+r_2$ sets of times. Then, let us denote $\hat{W}^{(n)}_{\pm k}(\vec u^{(i)})$, $\vec u^{(i)}=\{u^{(i)}_1,\cdots,u^{(i)}_n\}$, $i=1,\cdots,r_1$, \eqref{W+-}
 that act in the first $r_1$ sets of time variables $p_k^{(i)}$, $i=1,2,\ldots r_1$.
 \\\\
 Finally, we construct the following multicharacter partition function, with the use of the $W$-representation:
\begin{eqnarray}\label{tensor-r-wrep}
Z_{n,r_1+r_2}\{{\bf u}|\mathbf{ p}\}&=&
\exp\left(\sum_{k=1}^{\infty}
\dfrac{\prod_{i=1}^{r_1}\hat{W}^{(n)}_{-k}(\vec u^{(i)}) [p^{(i)}]\prod_{j=1}^{r_2} p_k^{(r_1+j)}}{k} \right)
\cdot 1 = \nonumber \\
&=&\sum_{R_1,\cdots,R_{r_1+r_2}}C_{R_1,\cdots,R_{r_1+r_2}}
\prod_{i=1}^{r_1} \left(\prod_{j=1}^n
\frac{S_{R_i}\{p_k^{(i)}=u^{(i)}_j\}}{S_{R_i}\{p_k^{(i)}=\delta_{k,1}\}} \right)
\prod_{s=1}^{r_1+r_2}S_{R_{s}}\{ p^{(s)}\},
\end{eqnarray}
where  ${\bf u}=\{\vec u^{(1)},\cdots,\vec u^{(r_1)}\}$,
$\mathbf{p}=\{p^{(1)},\cdots,p^{(r_1+r_2)}\}$,
$C_{R_1,\cdots,R_{r_1+r_2}}$ is the coefficient in the generalized Cauchy formula.
\\\\
Similarly we can construct the analog of the skew $\tau$-functions of \cite{MMMP}, which correspond to the positive branch of $W$-operators. Once again, take $r_1+r_2$ sets of times $p_k^{(i)}$ and $r_3$ sets of times $g_k^{(i)}$
\begin{eqnarray}\label{tensor-ext}
Z_{n,r_1+r_2+r_3}\{{\bf u}|\bf{p}|\bf{g}\}&=&
\exp\left(\sum_{m=1}^{\infty}
\frac{\prod_{i=1}^{r_1}\hat{W}^{(n)}_{m}(\vec u^{(i)}) \prod_{j=r_1+1}^{r_2}p^{(j)}_m}{m} \right)
\cdot
\exp\left(\sum_{k=1}^{\infty}\frac{\prod_{i=1}^{r_1} p_{k}^{(i)}
\prod_{j=1}^{r_3} g_{k}^{(j)}}{k} \right)=
\nonumber\\
&=&\sum_{Q_1,\cdots,Q_{r_1+r_2}\atop R_1,\cdots,R_{r_1+r_3}}
C_{Q_1,\cdots,Q_{r_1+r_2}}C_{R_1,\cdots,R_{r_1+r_3}}
\prod_{i=1}^{r_1}\prod_{j=1}^n
\frac{S_{R_i}\{p_l^{(i)}=u^{(i)}_j\}}
{S_{R_i}\{p_l^{(i)}=\delta_{l,1}\}} \times
\nonumber\\
&&\times\frac{S_{Q_i}\{p_l^{(i)}=\delta_{l,1}\}}
{S_{Q_i}\{p_l^{(i)}=u^{(i)}_j\}}
S_{R_i/Q_i}\{{ p}^{(i)}\}
\prod_{j=r_1+1}^{r_2}S_{Q_{j}}\{ p^{(j)}\}
\prod_{j=r_2+1}^{r_3}S_{R_{j}}\{g^{(j)}\},
\end{eqnarray}
where
$ \mathbf{g}=\{{ g}^{(1)},\cdots,{ g}^{(r_3)}\}$.
\\\\
We would like to look for integral representations for these two types of partition functions.
\subsection{Tensor models}
A well known representative of the family \eqref{tensor-r-wrep} with $r_1=r$, $r_2=0$ and $n=1$, $u_1^{(i)}=N^{(i)}$, is the partition function of the Gaussian tensor model \cite{Itoyama2020}, given by:
\begin{equation}
   \left\langle F(M,\bar{M}) \right\rangle =  \dfrac{\int  F(M,\bar{M}) \exp\left(\sum\limits_{ \left\{ a^{i}=1 \right\}}^{ \{N^{(i)} \}} M_{a^1, \ldots, a^r} \bar{M}^{a^1, \ldots, a^r} \right)dM d\bar M } {\int  \exp\left(\sum\limits_{ \left\{ a^{i}=1 \right\}}^{ \{N^{(i)} \}} M_{a^1, \ldots, a^r} \bar{M}^{a^1, \ldots, a^r} \right) dM d\bar M},
\end{equation}
where $dMd\bar M=\prod_{i=1}^{r}\prod_{a^i=1}^{N^{(i)}}dM_{a^1,\ldots, a^r}
d\bar M^{a^1,\ldots, a^r}$.
It has been represented in a $W$-operator like form using the automorphism $\hat{O}$:
\begin{eqnarray}\label{tensor-r}
Z_{r}\{{\bf p}\}
&=&\exp\left(\sum_{k=1}^{\infty}\frac{\prod_{i=1}^r\hat{O}_i(N^{(i)})p_k^{(i)}
{\hat{O}}^{-1}_i(N^{(i)})}
{k}\right)\cdot 1\nonumber\\
&=&
\sum_{R_1,\cdots,R_r}\prod_{i=1}^r S_{R_i}\{p^{(i)}\}
\left\langle S_{R_1,\cdots,R_r}\right\rangle\nonumber\\
&=&\sum_{R_1,\cdots,R_r}C_{R_1,\cdots,R_r}
\prod_{i=1}^r\dfrac{S_{R_i}\{p_l^{(i)}=N^{(i)}\}}
{S_{R_i}\{p_l^{(i)}=\delta_{l,1}\}}S_{R_i}\{p^{(i)}\}.
\end{eqnarray}
Here we used the expression for the avarages of generalized characters:
\begin{equation}
    S_{R_1, \ldots, R_r}(M, \bar{M})=\frac{1}{n !} \sum_{\sigma_1, \ldots, \sigma_r \in S_n} \psi_{R_1}\left(\sigma_1\right) \ldots \psi_{R_r}\left(\sigma_r\right) \cdot \mathcal{K}_{\sigma_1, \ldots, \sigma_r}^{(n)}
\end{equation}
with
\begin{equation}
    \mathcal{K}_{\sigma_1, \ldots, \sigma_r}^{(n)}=\sum_{\vec{a}^1=1}^{N^{(1)}} \ldots \sum_{\vec{a}^r=1}^{N^{(r)}}\left(\prod_{p=1}^n M_{a_p^1, \ldots a_p^r} \bar{M}^{a_{\sigma_1(p)}^1, \ldots, a_{\sigma_r(p)}^r}\right).
\end{equation}
These generalized characters enjoy an analog of \emph{superintegrability}, which was used in \eqref{tensor-r}:
\begin{equation}
    \left\langle S_{R_1,\ldots, R_r} \right\rangle = C_{R_1,\cdots,R_r}
\prod_{i=1}^r\frac{S_{R_i}\{p_l^{(i)}=N^{(i)}\}}
{S_{R_i}\{p_l^{(i)}=\delta_{l,1}\}}.
\end{equation}
\\\\
This representation has its own drawbacks. First of all, in the tensor model itself, the generalized characters form an overcomplete basis of operators. It's limit towards the complex model requires a reduction of the set of time variables as explained in \cite{Itoyama2020}. Furthermore, it allows the introduction of only "half" of time variables, which correspond to $r_2=0$ in \eqref{tensor-r-wrep}.
\\\\
For completeness we provide two more examples of tensor model partition functions. These examples correspond to choose the locus in such way that only a single set of independent time variables is left. Our results on $W$-operators provide an explicit $W$-representation for these tensor models.
\\\
\begin{itemize}
    \item  When $r_1=1$, $r_2=r-1$, $p_l^{(1)}=p_l$, $p_l^{(2)}=\cdots=p_l^{(r)}=\delta_{l,1}$
and $n=2$, $u^{(1)}_1=\textcolor{red}{N_1^{(1)}}$,
$u^{(1)}_2=\textcolor{green}{N_1^{(2)}}\textcolor{blue}{N_1^{(3)}}\cdots N_1^{(r)}$ in (\ref{tensor-r-wrep}), it reduces to the red tensor model \cite{ItoyamaJHEP2017,BKang2021}
\begin{equation}
    \begin{split}\label{red}
\textcolor{red}{Z_r\{p\}}
=&\int dM d\bar M \exp\left(-\Tr M\bar M+\sum_{m=1}^\infty {\frac{p_m}{m}}\textcolor{red}{\mathcal{T}_{(12\cdots m)\otimes id\otimes\cdots\otimes id}^{(m)}}
\right)\nonumber\\
=&\sum_R\frac{S_R\{p_l=\textcolor{red}{N_1^{(1)}}\}
S_R\{p_l=\textcolor{green}{N_1^{(2)}}\textcolor{blue}{N_1^{(3)}}
\cdots N_1^{(r)}\}
}{S_R\{p_l=\delta_{l,1}\}}S_R\{p\},
\end{split}
\end{equation}

where $\textcolor{red}{\mathcal{T}_{(12\cdots m)\otimes id\otimes\cdots\otimes id}^{(m)}}$ is the gauge invariant operator
\begin{equation}
    \begin{split}\label{redT}
\textcolor{red}{\mathcal{T}_{(12\cdots m)\otimes id\otimes\cdots\otimes id}^{(m)}}
=&\prod_{k=1}^{m-1}M_{\textcolor{red}{i}^{(k)},
{\textcolor{green}{j_1}^{(k)}},{\textcolor{blue}{j_2}^{(k)}},
\cdots,{j_{r-1}}^{(k)}}
\bar{M}^{\textcolor{red}{i}^{(k+1)},
{\textcolor{green}{j_1}^{(k)}},{\textcolor{blue}{j_2}^{(k)}},\cdots,
{j_{r-1}}^{(k)}}\times
\\
&\times
 M_{\textcolor{red}{i}^{(m)},{\textcolor{green}{j_1}^{(m)}},{\textcolor{blue}
 {j_2}^{(m)}},\cdots,{j_{r-1}}^{(m)}}
\bar{M}^{\textcolor{red}{i}^{(1)},{\textcolor{green}{j_1}^{(m)}},
{\textcolor{blue}{j_2}^{(m)}},\cdots,{j_{r-1}}^{(m)}}.
\end{split}
\end{equation}

 \item When $r_1=1$, $r_2=r-1$, $p_l^{(1)}=p_l$, $p_l^{(2)}=\frac{-1}{\textcolor{red}{N_1^{(1)}}\textcolor{green}
{N_1^{(2)}}\cdots N_1^{(r)}}\delta_{l,1}$,
$p_l^{(3)}=\cdots=p_l^{(r)}=\delta_{l,1}$
and $n=2$, $u^{(1)}_1=\textcolor{red}{N_1^{(1)}}$,
$u^{(1)}_2=-\textcolor{green}{N_1^{(2)}}\textcolor{blue}{N_1^{(3)}}\cdots N_1^{(r)}$ in (\ref{tensor-r-wrep}), it gives the character expansion of the fermionic red tensor model \cite{LYWang2021}
\begin{equation}
    \begin{split}
\textcolor{red}{\bar Z_r\{ p\}}
&=\frac{\int d\Psi d\bar \Psi \exp
\left(\textcolor{red}{N_1^{(1)}}\cdots N_1^{(r)}\Tr \bar\Psi\Psi +\sum_{{{m}}=1}^\infty
\frac{{{p_m}}}{{m}}\textcolor{red}{\mathcal{\bar T}_{(12\cdots m)\otimes id\otimes\cdots\otimes id}^{(m)}} \right)}{\int d\Psi d\bar\Psi\exp\left(\textcolor{red}{N_1^{(1)}}\cdots N_1^{(r)}\Tr\bar\Psi\Psi \right)}\\
&=\sum_R
\frac{1}{(\textcolor{red}{-N_1^{(1)}}\cdots N_1^{(r)})^{|R|}}
\frac{S_R\{p_l=\textcolor{red}{N_1^{(1)}}\}
S_R\{p_l=-\textcolor{green}{N_1^{(2)}}\textcolor{blue}{N_1^{(3)}}\cdots N_1^{(r)}\}}
{S_R\{p_l=\delta_{l,1}\}}S_R\{p_l\},
\end{split}
\end{equation}
where $\bar{\Psi}$ and $\Psi$ are the
fermionic tensors of rank~$r$ with one covariant and $r-1$ contravariant indices, the gauge invariant operator $\textcolor{red}{\mathcal{\bar T}_{(12\cdots m)\otimes id\otimes\cdots\otimes id}^{(m)}}$ is
defined by changing $M\rightarrow \Psi$ and $\bar M\rightarrow \bar \Psi$
in (\ref{redT}).
\end{itemize}

\subsection{Multimatrix model with multitrace potential}
Now suppose $r_1=r_2=r$ and denote "half" of the time variables as $p^{(i)}_l=\delta_{l,2}$, $p^{(r+i)}_l=\bar p^{(i)}_l$, $i=1,\cdots,r$. Put also  $n=1$, $u^{(i)}_1=N^{(i)}$ in (\ref{tensor-r-wrep}). The the following multi-matrix integral representation can be constructed:
\begin{eqnarray}\label{multigaussian}
&&\int \prod_{i=1}^{r}dX_i \exp\left(-\frac{1}{2}\sum_{i=1}^{r}\Tr X^2_i+\sum_{l=1}^{\infty}\frac{\prod_{i=1}^r g^{(i)}_k\Tr X_i^k}{k} \right)=
\nonumber\\
&=&\sum_{R_1,\cdots,R_{2r}}C_{R_1,\cdots,R_{2r}}
\prod_{i=1}^r\frac{S_{R_i}\{p_l^{(i)}=N^{(i)}\}}
{S_{R_i}\{p_l^{(i)}=\delta_{l,1}\}}S_{R_i}\{p_l^{(i)}=\delta_{l,2}\}
S_{R_{r+i}}\{g^{(i)}\},
\end{eqnarray}
where $X_i$ is an $N^{(i)}\times N^{(i)}$ matrix.
\\\\
Just as in the Gaussian one matrix model this partition function is limited to putting a subset of times variables to $\delta_{k,2}$. It seems the construction in \cite{MMMP,Alexandrov:2022whk} could be used to include a generic set of the second "half" of times and even further to include skew partition functions. Indeed, introducing integration over new matrices $Y_i$ for each $i=1,\ldots,r$ we couple them to $X_i$ in the following way:
\begin{equation}
\begin{split}
    Z=&\int \prod_{i=1}^{r} dX_i dY_i \exp\left( \sum_{i=1}^r\Tr\, X_i Y_i +\sum_{k=1}^{\infty}\frac{\prod_{i=1}^r g^{(i)}_k\Tr X_i^k}{k} + \sum_{i=1}^r\sum_{k=1}^{\infty} \dfrac{\bar{p}^{(i)}_k \Tr Y_i^k}{k} \right) =\\
    =&\sum_{R_1,\ldots R_{2r}} C_{R_1,\cdots,R_{2r}} \sum_{Q_1,\ldots Q_r} \prod_{i=1}^{r}  S_{R_{r+i}} \{g^{(i)}\} \left( \prod_{i=1}^{r} S_{Q_{i}}\{\bar{p}^{(i)}\} \Big\langle S_{R_i}\left\{X_i \right\} S_{Q_i}\left\{Y_i \right\} \Big\rangle \right) = \\
    &= \sum_{R_1,\cdots,R_{2r}}C_{R_1,\cdots,R_{2r}}
\prod_{i=1}^r\frac{S_{R_i}\{p_l^{(i)}=N^{(i)}\}}
{S_{R_i}\{p_l^{(i)}=\delta_{l,1}\}}S_{R_i}\{\bar{p}^{(i)}\}
S_{R_{r+i}}\{g^{(i)}\}.
\end{split}
\end{equation}
To further incorporate the external matrices, notice that in \cite{MMMP} it was achieved by introducing a shift in the $X$ variables. Hence we propose the following model for the skew multicharacter expansion:
\begin{equation}
\begin{split}
    Z=&\int \prod_{i=1}^{r} dX_i dY_i \exp\left( \sum_{i=1}^r \Tr\, \left(X_i - \Lambda_i \right) Y_i+\sum_{k=1}^{\infty}\frac{\prod_{i=1}^r g^{(i)}_k\Tr X_i^k}{k} + \sum_{k=1}^{\infty} \dfrac{\prod_{i=1}^r\bar{p}^{(i)}_k \Tr Y_i^k}{k} \right) =\\
    =&\sum_{Q_1,\cdots,Q_{2r}\atop R_1,\cdots,R_{2r}}
C_{Q_1,\cdots,Q_{2r}}C_{R_1,\cdots,R_{2r}}
\prod_{i=1}^{r}
\dfrac{S_{Q_i}\{p_k^{(i)}=N^{(i)}\} }{S_{Q_i}\{p_k^{(i)}=\delta_{k,1} \}}
S_{R_i/Q_i}\{{p}^{(i)}\}
\prod_{j=1}^{r}S_{Q_{r+j}}\{ \bar{p}^{(j)}\}
\prod_{j=1}^{r}S_{R_{r+j}}\{g^{(j)}\}.
\end{split}
\end{equation}
Note the different entrance of the couplings $\bar{p}$, which is necessary to correctly reproduce the structure of the generalized Clebsh-Gordan coefficients.
\section{$\beta$-deformation}\label{sec:beta}

In this section we would like to demonstrate that the $\beta$-deformation of all of the $W$-representations in section above are straightforward. This suggests to look for their integral representations as well \cite{Mironov:2023mve}. Here we focus on the operators and leave the subject of $\beta$-deformed integrals for the future.

\paragraph{Operators}

In order to proceed, we define the $\beta$-deformed diagonal operator:
\begin{eqnarray}\label{w02beta}
\hat{\mathcal{W}}_0(u)
=\frac{1}{2}\sum_{k,l=1}^{\infty}\left(\beta(k+l)p_k p_l\frac{\partial}{\partial p_{k+l}}
+klp_{k+l}\frac{\partial ^2}{\partial p_k\partial p_l} \right)
+\frac{1}{2}\sum_{k=1}^{\infty}\left((1-\beta)(k-1)+2\beta u \right)kp_k\frac{\partial}{\partial p_k},
\end{eqnarray}
where $u$ is an arbitrary parameter. Just as in the $\beta=1$ case, it can be used to construct commutative families of operators:
\begin{eqnarray}\label{w-knku-beta}
\hat{\mathcal{W}}_{-k}^{(n)}(\vec{u})
=\frac{1}{(k-1)!}\ad^{k-1}_{\hat{\mathcal{W}}_{-1}^{(n+1)}(\vec{u})} \hat{\mathcal{W}}_{-1}^{(n)}(\vec{u}),
\end{eqnarray}
where $\vec{u}=(u_1,u_2,\cdots,u_n)$ and $\hat{\mathcal{W}}_{-1}^{(n)}(\vec{u})=\left[\hat{\mathcal{W}}_0(u_n),
\left[\hat{\mathcal{W}}_0(u_{n-1}),\cdots,\left[\hat{\mathcal{W}}_0(u_1),p_1 \right]\cdots \right]\right].$
\\\\
The operators act on the corresponding deformation of characters - the Jack polynomials:
\begin{eqnarray}\label{w-kk-jack}
\hat{\mathcal{W}}^{(n)}_{-k}(\vec u)J_R
=\sum_{P} \prod_{(i,j) \in P/R } \left(\prod_{s=1}^{n}
 (\beta(u_s-i+1)+j-1) \right)
\frac{\Big\langle p_k J_R \Big| J_{P} \Big\rangle_\beta}{\Big\langle  J_{P} \Big| J_{P} \Big\rangle_\beta} J_{P},
\end{eqnarray}
where we introduced the $\beta$-deformed norm:
\begin{equation}
    \Big\langle  J_R \Big| J_{P} \Big\rangle_\beta=
    \prod_{(i,j)\in R}
    \frac{R_i-j+1+\beta(R'_j-i)}{R_i-j+\beta(R'_j-i+1)}\delta_{R,P},
\end{equation}
with $R'$ the conjugate partition of $R$.
\paragraph{Partition functions}

In terms of $\hat{\mathcal{W}}^{(n)}_{-k}(\vec u)$, one immediately construct the $\beta$-deformed partition functions of any type of those discussed in section \ref{sec:multichar}. Beginning with the deformation of the Hurwitz $\tau$-functions:
\begin{eqnarray} \label{Z2n}
\mathcal{Z}_{n}\{\vec u|p,\bar{p}\}&=&
\exp\left( \beta\sum_{k=1}^{\infty} \hat{\mathcal{W}}_{-k}^{(n)}(\vec{u})\frac{{\bar p}_k}{k} \right)\cdot1\nonumber\\
&=&\sum_R\prod_{i=1}^n\frac{J_R\{p_k=u_i\}}{J_R\{p_k=\beta^{-1}\delta_{k,1}\}}
\frac{J_R\{p\}J_R\{\bar{p}\}}{\Big\langle  J_R \Big| J_R \Big\rangle_\beta}.
\end{eqnarray}
Particular reduction of these expansions involves several cases well known in the literature such as:
\begin{itemize}
    \item When $n=2$, $u_1=N$, $u_2=N+\beta^{-1}\nu+\beta^{-1}-1$ and $\bar p_k=\beta^{-1}a_1^{-1}\delta_{k,1}$ in (\ref{Z2n}), it gives the supposed $\beta$-deformed of the complex (square) matrix model with a logarithmic insertion of \cite{Cassia,Bawane:2022cpd}:
\begin{eqnarray}\label{logpot}
{\mathcal Z}\{p\}&=&\prod_{i=1}^N\int_{0}^{+\infty} dz_i \Delta(z)^{2\beta}
e^{\sum_{i=1}^{N}\sum_{k=1}^{\infty}\beta\frac{p_k}{k}z_i^k
-\sum_{i=1}^N(-a_1z_i+\nu {\rm log}z_i)}\nonumber\\
&=&\sum_{R}\frac{J_{R}\{p_k=N\}
J_{R}\{p_k=N+\beta^{-1}\nu+\beta^{-1}-1\}}{J_{R}\{p_k=a_1\beta^{-1}\delta_{k,1}\}
\Big\langle  J_R \Big| J_R \Big\rangle_\beta }J_{R}\{p\}.
\end{eqnarray}
It is obvious that when $\nu=0$, (\ref{logpot}) is the $\beta$-deformed complex matrix model \cite{Cheny}.

\item When $n=1$, $u_1=N$ and $\bar p_k=\beta^{-1}a_2^{-1}(\delta_{k,2}-a_1\delta_{k,1})$ in (\ref{Z2n}), it gives the $\beta$-deformed eigenvalue whose potential is the Gaussian potential with a linear term \cite{Cassia,Bawane:2022cpd}:
\begin{eqnarray} \label{Z21-m-beta}
\bar {\mathcal{Z}}\{p\}
&=&\prod_{i=1}^N\int_{-\infty}^{+\infty} dz_i\Delta(z)^{2\beta}
e^{\sum_{i=1}^{N}\sum_{k=1}^{\infty}\beta\frac{ p_k}{k}z_i^k-
\sum_{i=1}^N(\frac{a_2}{2}z_i^2+a_1z_i)}\nonumber\\
&=&\sum_{R}
\frac{J_{R}\{p_k=N\}J_{R}\{p_k=\beta^{-1}(a_2\delta_{k,2}-a_1\delta_{k,1})\}}
{J_{R}\{p_k=a_2\beta^{-1}\delta_{k,1}\}
\Big\langle  J_R \Big| J_R \Big\rangle_\beta}
J_{R}\{p\}.
\end{eqnarray}

\end{itemize}

However these two examples are reductions are a very specific locus. Instead it would be interesting to construct a $\beta$-deformed analog of the multimatrix model \cite{MMMP,Alexandrov:2022whk}, which was done in \cite{Mironov:2023mve}.
Finishing with the negative branch partition functions, the multi-character expansions clearly also have a counterpart:
\begin{eqnarray}\label{deftensor-r-beta}
\mathcal{Z}_{n,r_1+r_2}\{{\bf u}|{\bf p}\}
&=&\exp\left(\sum_{k=1}^{\infty}
\frac{\prod_{i=1}^{r_1}\beta\hat{\mathcal{W}}^{(n)}_{-k}(\vec u^{(i)}) \prod_{j=1}^{r_2}\beta p^{(r_1+j)}_k}{k} \right)
\cdot 1\nonumber\\
&=&\sum_{R_1,\cdots,R_{r_1+r_2}}\mathcal{C}_{R_1,\cdots,R_{r_1+r_2}}
\prod_{i=1}^{r_1}\prod_{j=1}^n
\frac{J_{R_i}\{p_l^{(i)}=u^{(i)}_j\}}{J_{R_i}\{p_l^{(i)}=\beta^{-1}\delta_{l,1}\}}
\prod_{s=1}^{r_1+r_2}J_{R_{s}}\{ p^{(s)}\},
\end{eqnarray}
where the coefficients $\mathcal{C}_{R_1,R_2,\ldots, R_n}$ are the straightforward deformation of the generalized Chebsh-Gordan coefficients which are determined by the expansion
\begin{eqnarray}\label{generalized-Cauchy-beta}
\exp\left(\sum_{k=1}^{\infty}\frac{\prod_{i=1}^r\beta p^{(i)}_{k}}{k} \right)
=\sum_{R_1,\cdots,R_r}
\mathcal{C}_{R_1,\cdots,R_r}\prod_{i=1}^rJ_{R_i}\{p^{(i)}\}.
\end{eqnarray}
\\\\
The same goes for the positive branch, where we use the operators (\ref{w02beta}) and $\frac{\partial}{\partial p_1}$,
to obtain
\begin{eqnarray}\label{wknkui-beta}
\hat{\mathcal{W}}_{k}^{(n)}(\vec{u})=
\frac{(-1)^{k-1}}{(k-1)!} \ad^{k-1}_{\hat{\mathcal{W}}_{1}^{(n+1)}(\vec{u})} \mathcal{W}_{1}^{(n)}(\vec{u}),
\end{eqnarray}
where $\hat{\mathcal{W}}_{1}^{(n)}(\vec{u})=
\left[\cdots \left[\left[\beta^{-1}\frac{\partial}{\partial p_1},\hat{\mathcal{W}}_{0}(u_1)\right],\hat{\mathcal{W}}_{0}(u_2)\right],\cdots,\hat{\mathcal{W}}_{0}(u_n)\right].
$
\\\\
Their action on Jack functions is given by:
\begin{eqnarray}\label{w-kk-jack}
	\hat{\mathcal{W}}^{(n)}_{k}(\vec u)J_R
	=\sum_{P} \prod_{(i,j) \in R/P } \left(\prod_{s=1}^{n}
	(\beta(u_s-i+1)+j-1) \right)
	\frac{\Big\langle \dfrac{k}{\beta} \dfrac{\partial}{\partial p_k} J_R \Big| J_{P} \Big\rangle_\beta}{\Big\langle  J_{P} \Big| J_{P} \Big\rangle_\beta} J_{P},
\end{eqnarray}
Then the skew Jack expansion is given by:
\begin{eqnarray}\label{z2n-ex-beta}
\mathcal{Z}_{n} \{\vec u|p,\bar{p}|g\}&=&
\exp\left({\beta\sum_{k=1}^{\infty}\hat{\mathcal{W}}_{k}^{(n)}(\vec{u})\frac{\bar p _k}{k}} \right) \cdot \exp\left({\beta\sum_{k=1}^{\infty} \frac{p_k g_k}{k}} \right)
\nonumber\\
&=&\sum_{R,Q}\, \prod_{i=1}^n\frac{J_R\{p_k=u_i\}J_Q\{p_k=\beta^{-1}\delta_{k,1}\}}
{J_R\{p_k=\beta^{-1}\delta_{k,1}\}J_Q\{p_k=u_i\}}
\frac{J_{R/Q}\{\bar p\}J_Q\{p\}J_R\{g\}}
{\Big\langle  J_R \Big| J_R \Big\rangle_\beta}
\end{eqnarray}
for two sets of times and the multi-character generalisation is given by:
\begin{eqnarray}\label{deftensor-r-beta-ext}
\mathcal{Z}_{n,r_1+r_2+r_3}\{{\bf u}|{\bf p}|{\bf g}\}&=&
\exp\left(\sum_{k=1}^{\infty}
\frac{\prod_{i=1}^{r_1}\beta\hat{\mathcal{W}}^{(n)}_{k}(\vec u^{(i)}){ \prod_{j=1}^{r_2}\beta p_k^{(r_1+j)}}}{k} \right) \cdot
\exp\left(\sum_{k=1}^{\infty}\frac{\prod_{i=1}^{r_1} \beta p_{k}^{(i)}
\prod_{j=1}^{r_3}\beta g_{k}^{(j)} }{k} \right)\nonumber\\
&=&\sum_{Q_1,\cdots,Q_{r_1+r_2}\atop R_1,\cdots,R_{r_1+r_3}}
\mathcal{C}_{Q_1,\cdots,Q_{r_1+r_2}}\mathcal{C}_{R_1,\cdots,R_{r_1+r_3}}
\prod_{i=1}^{r_1}\prod_{j=1}^n
\frac{J_{R_i}\{p_l^{(i)}=u^{(i)}_j\}}
{J_{R_i}\{p_l^{(i)}=\delta_{l,1}\}}
\nonumber\\
&&\times \frac{J_{Q_i}\{p_l^{(i)}=\delta_{l,1}\}}
{J_{Q_i}\{p_l^{(i)}=u^{(i)}_j\}}
\frac{J_{R_i/Q_i}\{{ p}^{(i)}\}}{\Big\langle J_{Q_i}|J_{Q_i}\Big\rangle_{\beta}}
\prod_{s=1}^{r_2}J_{Q_{r_1+s}}\{ p^{(r_1+s)}\}
\prod_{t=1}^{r_3}J_{R_{r_1+t}}\{g^{(t)}\}.
\end{eqnarray}
Here the skew Jack polynomials are  $J_{R/Q}=\sum\limits_{P}\frac{\Big\langle  J_R \Big| J_Q J_P \Big\rangle_\beta}{\Big\langle  J_Q \Big| J_Q \Big\rangle_\beta\Big\langle  J_P \Big| J_P \Big\rangle_\beta}J_P$ are the skew Jack polynomials.

\section *{Acknowledgments}
We are grateful to A. Morozov and A. Mironov for helpful comments. This work is supported by the grant of the Foundation for the Advancement of Theoretical Physics ``BASIS" and by the joint grant 21-51-46010-ST-a and the National Natural Science Foundation of China (No. 12205368).


\begin{thebibliography}{21}
\bibitem{Itoyama2020}
H. Itoyama, A. Mironov and A. Morozov,
Complete solution to Gaussian tensor model and its integrable properties,
Phys. Lett. B {\bf 802} (2020) 135237 [arXiv:1910.03261].

\bibitem{MMMP}
A. Mironov, V. Mishnyakov, A. Morozov, A. Popolitov, R. Wang and W.Z. Zhao, Interpolating Matrix Models for WLZZ series, arXiv:2301.04107.

\bibitem{wangruiliufan}
R. Wang, F. Liu, C.H. Zhang and W.Z. Zhao, Superintegrability for ($\beta$-deformed) partition function hierarchies with $W$-representations,
Eur. Phys. J. C {\bf 82} (2022) 902 [arXiv:2206.13038].

\bibitem{Mishnyakov:2022bkg}
V.~Mishnyakov and A.~Oreshina,
``Superintegrability in $\beta $-deformed Gaussian Hermitian matrix model from W-operators,''
Eur. Phys. J. C \textbf{82} (2022) no.6, 548
doi:10.1140/epjc/s10052-022-10466-y
[arXiv:2203.15675 [hep-th]].


\bibitem{Morozov:2019gbt}
A.~Morozov,
``On $W$-representations of $\beta$- and $q,t$-deformed matrix models,''
Phys. Lett. B \textbf{792} (2019), 205-213
doi:10.1016/j.physletb.2019.03.047
[arXiv:1901.02811 [hep-th]].


\bibitem{Bawane:2022cpd}
A.~Bawane, P.~Karimi and P.~Su\l{}kowski,
``Proving superintegrability in $\beta$-deformed eigenvalue models,''
SciPost Phys. \textbf{13} (2022) no.3, 069
doi:10.21468/SciPostPhys.13.3.069
[arXiv:2206.14763 [hep-th]].

\bibitem{1405}
A. Alexandrov, A. Mironov, A. Morozov and S. Natanzon, On KP-integrable Hurwitz functions,
J. High Energy Phys. {\bf 11} (2014) 080 [arXiv:1405.1395].

\bibitem{Itoyama1909}
 H. Itoyama, A. Mironov and A. Morozov, Tensorial generalization of characters, J. High Energy Phys. {\bf 12} (2019) 127 [arXiv:1909.06921].


\bibitem{BKang2021}
B. Kang, L.Y. Wang, K. Wu, J. Yang and W.Z. Zhao, $W$-representation of rainbow tensor model, J. High Energy Phys. {\bf 05} (2021) 228 [arXiv:2104.01332].

\bibitem{ItoyamaJHEP2017}
H. Itoyama, A. Mironov and A. Morozov, Ward identities and combinatorics of rainbow tensor models,
J. High Energy Phys.  {\bf 06} (2017) 115 [arXiv:1704.08648].

\bibitem{LYWang2021}
L.Y. Wang, R. Wang, K. Wu and W.Z. Zhao, $W$-representations of the fermionic matrix and Aristotelian tensor models,
Nucl. Phys. B {\bf 973} (2021) 115612 [arXiv:2110.14269].

\bibitem{Alexandrov:2022whk}
A.~Alexandrov,
``On $W$-operators and superintegrability for dessins d'enfant,''
[arXiv:2212.10952 [hep-th]].

\bibitem{Mironov:2023mve}
A.~Mironov, V.~Mishnyakov, A.~Morozov, A.~Popolitov and W.~Z.~Zhao,
[arXiv:2301.11877 [hep-th]].

\bibitem{Cassia}
L. Cassia, R. Lodin and M. Zabzine, On matrix models and their $q$-deformations, J. High Energy Phys. {\bf 10} (2020) 126 [arXiv:2007.10354].


\bibitem{Cheny}
Y. Chen, B. Kang, M.L. Li, L.F. Wang and C.H. Zhang, Correlators in the $\beta$-deformed Gaussian Hermitian and complex matrix models,
Int. J. Mod. Phys. A {\bf34} (2019) 1950221.



\end{thebibliography}
\end{document}